\documentclass[aps,prl,twocolumn,preprintnumbers,10pt,showpacs,nohyper]{revtex4}
\usepackage{amsmath,bm}
\usepackage{multirow}

\usepackage{graphicx}
\usepackage{xspace}
\usepackage{epsfig}
\usepackage{color}

\newcommand{\NNLOJET}{NNLO\protect\scalebox{0.8}{JET}\xspace}

\def\ys{$|y^*|$\xspace}

\allowdisplaybreaks

\begin{document}

\preprint{IPPP/17/45, ZU-TH 13/17, MPP-2017-107}

\title{Precise predictions for dijet production at the LHC}

\author{J. Currie$^a$, A. Gehrmann-De Ridder$^{b,c}$, T. Gehrmann$^{c}$, E.W.N. Glover$^a$, A. Huss$^{b}$, J. Pires$^d$}
\affiliation{$^a$ Institute for Particle Physics 
Phenomenology, University of Durham, Durham DH1 3LE, UK\\
$^b$ Institute for Theoretical Physics, ETH, CH-8093 Z\"urich, Switzerland 
\\
$^c$ Department of Physics, Universit\"at Z\"urich, Winterthurerstrasse 190, CH-8057
Z\"urich, Switzerland 
\\
$^d$ Max-Planck-Institut f\"ur Physik,
F\"ohringer Ring 6,
D-80805 Munich, Germany}


\begin{abstract}

We present the calculation of dijet production, doubly-differential in dijet mass, $m_{jj}$ and rapidity difference, \ys, 
at leading colour in all partonic channels at next-to-next-to-leading order (NNLO) in perturbative QCD.
We consider the long-standing problems associated with scale choice for dijet production at next-to-leading order (NLO) and
investigate the impact of including the NNLO contribution. We find that the NNLO theory provides reliable predictions,
even when using scale choices which display pathological behaviour at NLO. We choose the dijet invariant mass as the
theoretical scale on the grounds of perturbative convergence and residual scale variation
and compare the predictions to the ATLAS 7 TeV 4.5~fb$^{-1}$ data.

\end{abstract}

\maketitle

The production of jets in the final-state is one of the most frequently occurring reactions at hadron colliders, such as the LHC.
When at least two jets are produced, the two jets leading in transverse momentum, $p_{T}$, constitute a 
dijet system. Such systems are a powerful tool when searching for physics beyond the Standard Model
by ``bump hunting'' in the dijet mass spectrum~\cite{dijet1, dijet2, dijet3, dijet4} or testing the QCD running coupling to very large momentum 
transfer~\cite{cms32, atlas32}.
Even in the case of no new physics being found, dijet observables offer a win-win scenario as they can provide
valuable information on important Standard Model parameters such as the strong coupling, $\alpha_{s}$, and
the Parton Distribution Functions (PDFs).

To fully exploit the wealth of available data it is important to have a reliable and accurate theoretical prediction. The dijet
observables considered in this letter are currently known to NLO accuracy in perturbative QCD~\cite{eks,jetrad,powheg2j,nlojet1,meks} 
and electroweak effects~\cite{eweak, eweak2,eweak3}.
Although the NLO corrections give an improvement on the LO prediction, there remains significant theoretical
uncertainty associated with the NLO calculation. It is well known that the parametric choice of scales for renormalization, $\mu_{R}$,
and factorization, $\mu_{F}$, 
has a big impact on the predictions at NLO and, for this reason, the dijet data is regularly excluded from global PDF fits.
To improve the theoretical description of dijet observables and make a meaningful comparison to data
 it is therefore necessary to calculate dijet production to  NNLO accuracy. The NNLO correction to jet production was first discussed in the context
 of the single jet inclusive cross section~\cite{incljet, incljetpt} and in this letter we report, for the first time, the NNLO corrections to dijet production.

At hadron colliders, jets are reconstructed by applying a jet algorithm~\cite{jetsgavin} and ordered in transverse momentum. 
The LHC experiments have measured 
dijet events~\cite{cmsdijet, atlasdijet} at 7~TeV as distributions in the dijet invariant mass, $m_{jj}$,
\begin{eqnarray}
m_{jj}^2&=&(p_{j_{1}}+p_{j_{2}})^{2},
\end{eqnarray}
where $p_{j_{1,2}}$ are the four-momenta of the two leading jets in an event satisfying the fiducial cuts, and the rapidity difference, \ys, where,
\begin{eqnarray}
y^* &=& \frac{1}{2} (y_{j_{1}}-y_{j_{2}}),
\end{eqnarray}
and $y_{j_{1,2}}$ are the rapidities of the two leading jets. For two exactly balanced (back-to-back) jets,
the invariant mass is related to the transverse momentum, $p_{T}$, and $y^*$ variables by the simple relation,
\begin{eqnarray}
m_{jj}&=&2p_{T}\cosh(y^*).\label{eq:mjjdef}
\end{eqnarray}
This relation always holds at leading order (LO) but is modified at NLO and NNLO by the presence of 
additional real radiation contributions. It is 
evident from Eq.~(\ref{eq:mjjdef}) that a minimum $p_{T}$ cut translates to a minimum accessible value of $m_{jj}$
which increases with \ys, such that in bins of large \ys only large values of $m_{jj}$ are experimentally accessible.

 \begin{figure*}[t]
  \centering
    \includegraphics[width=0.45\textwidth]{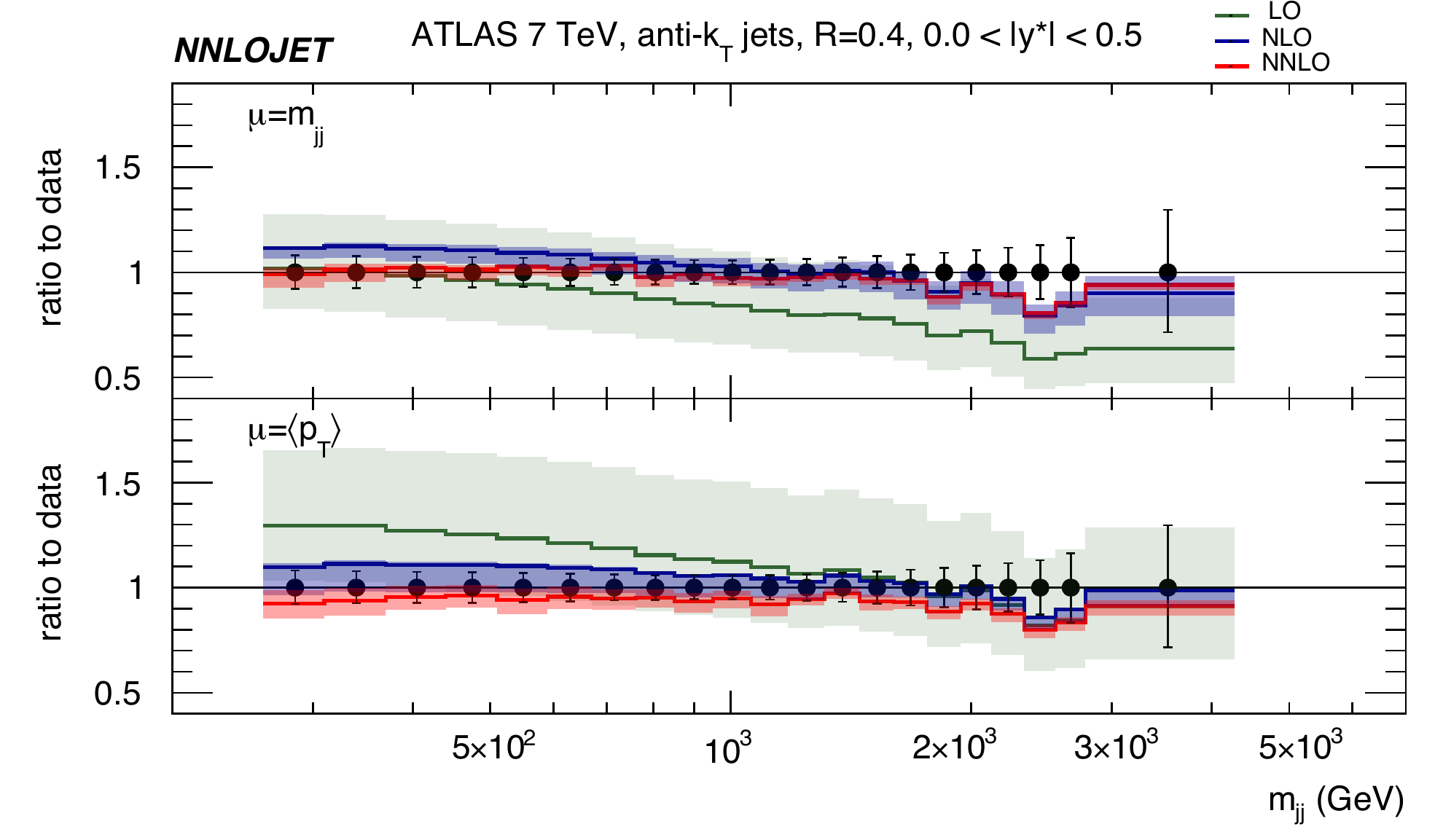}\hspace{1cm}
    \includegraphics[width=0.45\textwidth]{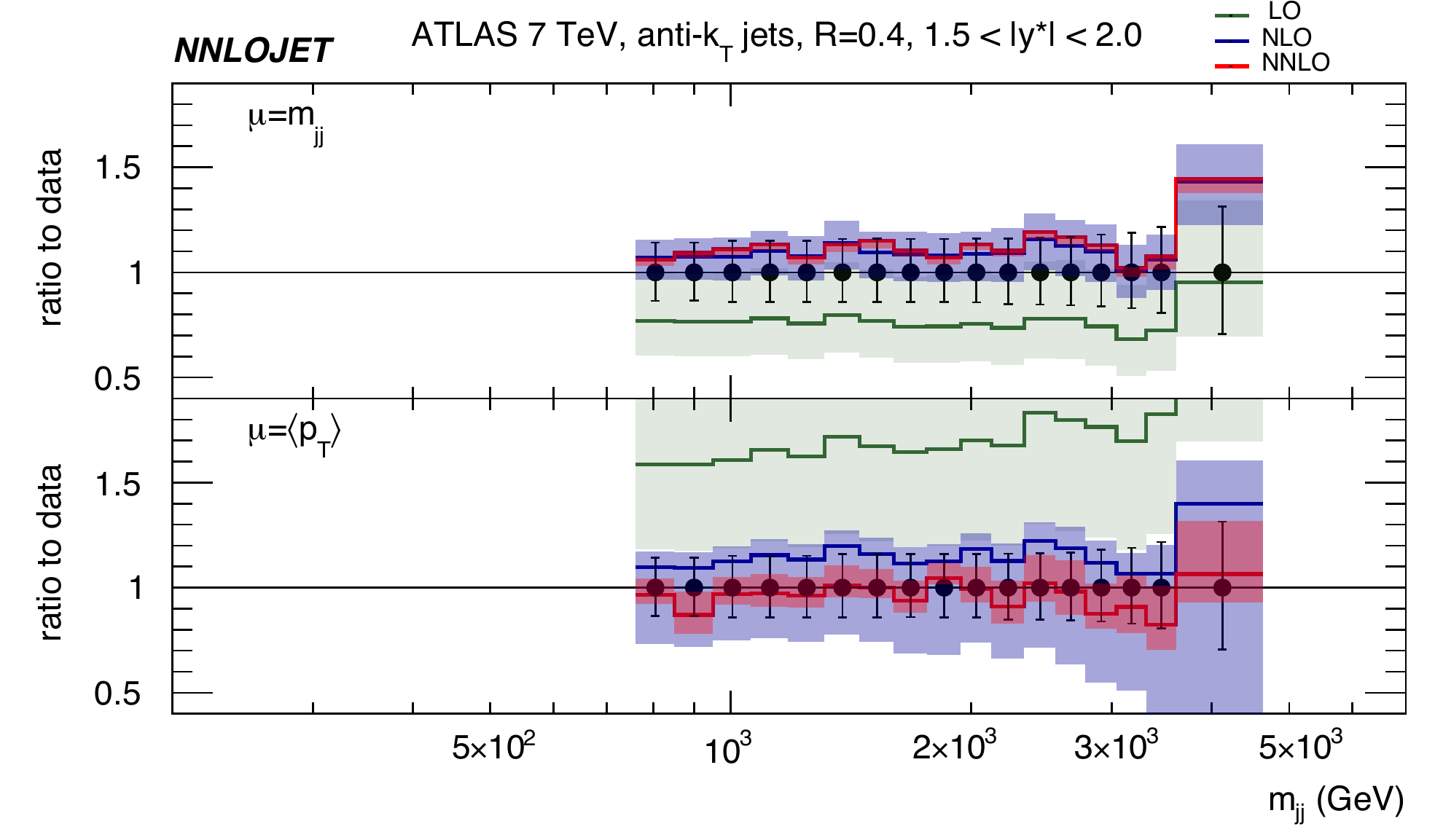}
  \caption{Ratio of theory predictions to data for $0.0<|y^*|<0.5$ (left) and $1.5<|y^*|<2.0$ (right) for the scale choices $\mu=m_{jj}$ (top) and $\mu=\langle p_{T}\rangle$ (bottom) at LO (green), NLO (blue) and NNLO (red). Scale bands represent variation of the cross section by varying the scales independently by factors of 2 and 0.5.}
  \label{fig:scales}
\end{figure*}

The longitudinal momentum fractions of the incoming partons
can, for back-to-back jets, be written in terms of the final-state jet parameters using momentum conservation,
\begin{eqnarray}
x_{1}&=&\frac{1}{2}x_{T}~(\mathrm{e}^{+y_{j_{1}}}+\mathrm{e}^{+y_{j_{2}}})=x_{T}~\mathrm{e}^{+\bar{y}}\cosh(y^*),\nonumber\\
x_{2}&=&\frac{1}{2}x_{T}~(\mathrm{e}^{-y_{j_{1}}}+\mathrm{e}^{-y_{j_{2}}})=x_{T}~\mathrm{e}^{-\bar{y}}\cosh(y^*),\label{eq:x12}
\end{eqnarray}
where $x_{T}=2p_{T}/\sqrt{s}$ and $\bar{y}=\frac{1}{2}(y_{1}+y_{2})$ is the rapidity of the dijet system in the lab frame. 
From Eq.~(\ref{eq:x12}) it is clear that for small values of $\bar{y}$ the dijet data probe
the configuration $x_{1}\approx x_{2}$, with the $x$-values determined by the $p_{T}$ of the jets. 
For large rapidities the data probes the scattering
of a high-$x$ parton off a low-$x$ parton. By binning the data in \ys, these configurations are smeared out
across the distribution and so a single bin in \ys will contain a wide range of possible $x$-values. This is in
contrast to binning in the maximum rapidity $y_{\mathrm{max}}$, as was done for dijet studies at the D$\O$ experiment~\cite{d0dijet},
or the triply differential distribution in $p_{T_{1}}$, $y_{1}$ and $y_{2}$ (or alternatively, average jet $p_{T}$, \ys and $|\bar{y}|$)~\cite{triplediff,cmstriplediff},
which would provide more specific information on the $x$-values probed.

The data sample we compare to is the ATLAS 7 TeV 4.5~fb$^{-1}$ 2011 data~\cite{atlasdijet}. 
This constitutes the recording of all events with at least two jets reconstructed in the rapidity range $|y| < 3.0$ using the anti-$k_t$ algorithm with R=0.4 such that the leading and subleading jets satisfy a minimum $p_{T}$ cut of 100 GeV and 50 GeV respectively.

As detailed in~\cite{incljet}, we include the leading colour NNLO corrections in all partonic sub-processes.
 The calculation is performed in the \NNLOJET framework, which employs the antenna subtraction
 method~\cite{antenna1, antenna2} to remove all unphysical infrared singularities from the matrix elements~\cite{real, 1-loop, 2-loop}. We use the 
 MMHT2014 NNLO parton distribution functions~\cite{mmht14} with 
 $\alpha_{s}(M_{Z})=0.118$ for all predictions at LO, NLO and NNLO to emphasize the role of the perturbative corrections at each successive order.

At any given fixed order in perturbation theory, the predictions retain some dependence on the unphysical renormalization and factorization scales.
The natural physical scale for dijet production is the dijet invariant mass, $\mu=m_{jj}$, which has not been widely used in dijet studies to date. 
Another scale, which was used at D$\O$~\cite{d0dijet} and is currently used by CMS~\cite{cmsdijet}
is the average $p_{T}$ of the two leading jets, $\mu=\langle p_{T}\rangle=\frac{1}{2}(p_{T_{1}}+p_{T_{2}})$. 

    \begin{figure}[b]
  \centering
    \includegraphics[width=0.5\textwidth]{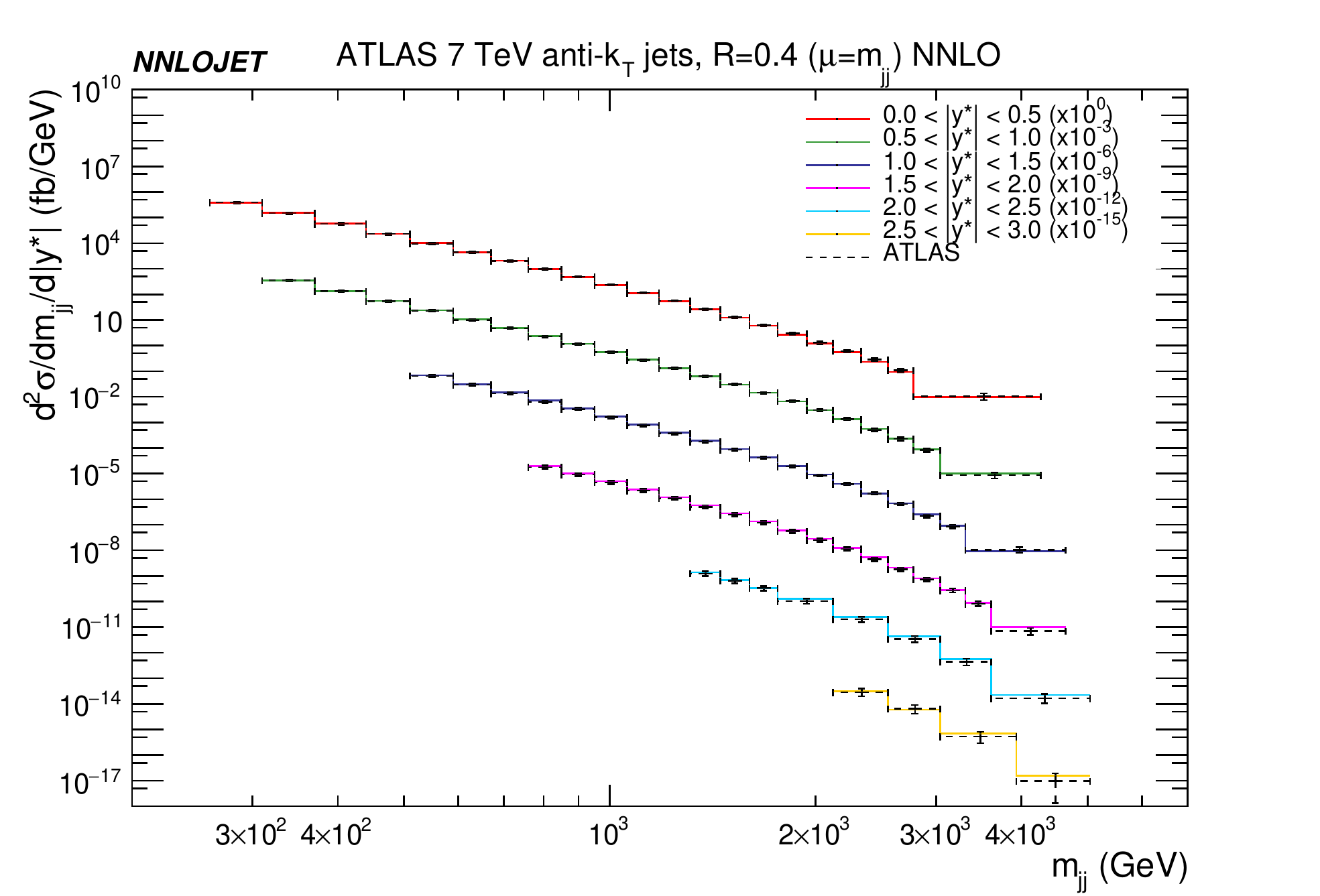}
  \caption{The dijet cross section as a function of invariant mass, $m_{jj}$, for the six bins of \ys, compared to ATLAS
  7~TeV 4.5~fb$^{-1}$ data.}
  \label{fig:xsec}
\end{figure}

  In Fig.~\ref{fig:scales} we show the predictions at LO, NLO and NNLO for these two scale choices at small and large \ys.
  For small \ys, both scale choices provide reasonable predictions with largely overlapping scale bands, reduced scale variation at each perturbative order,
   convergence of the perturbative series and good description of the data. For the larger \ys bin we see significant differences in
   the behaviour of the predictions for the two scales. For the $\mu=m_{jj}$ scale choice, the behaviour is qualitatively similar to what is seen at small \ys;
in contrast, the 
NLO prediction with $\mu=\langle p_{T}\rangle$ falls well away from the LO prediction and is even outside the LO scale band. For this scale choice, the NLO contribution
   induces a large negative correction, which brings the central value in line with the data but with a residual
   scale uncertainty of up to 100\%. Indeed for \ys$>$2.0 the scale band for $\mu=\langle p_{T}\rangle$ widens further and even 
   includes negative values of the cross section. These issues are resolved by the inclusion of the NNLO contribution 
   such that the NNLO prediction is positive across the entire phase space and provides a good description
   of the data.
  With the issue of unphysical predictions
  resolved, we are free to make a scale choice based upon
  more refined qualities such as perturbative convergence and residual scale variation. On this basis we choose the theoretical scale $\mu=m_{jj}$
  and present detailed results using this scale choice throughout the rest of this letter.

 \begin{figure}[t]
  \centering
    \includegraphics[width=0.45\textwidth]{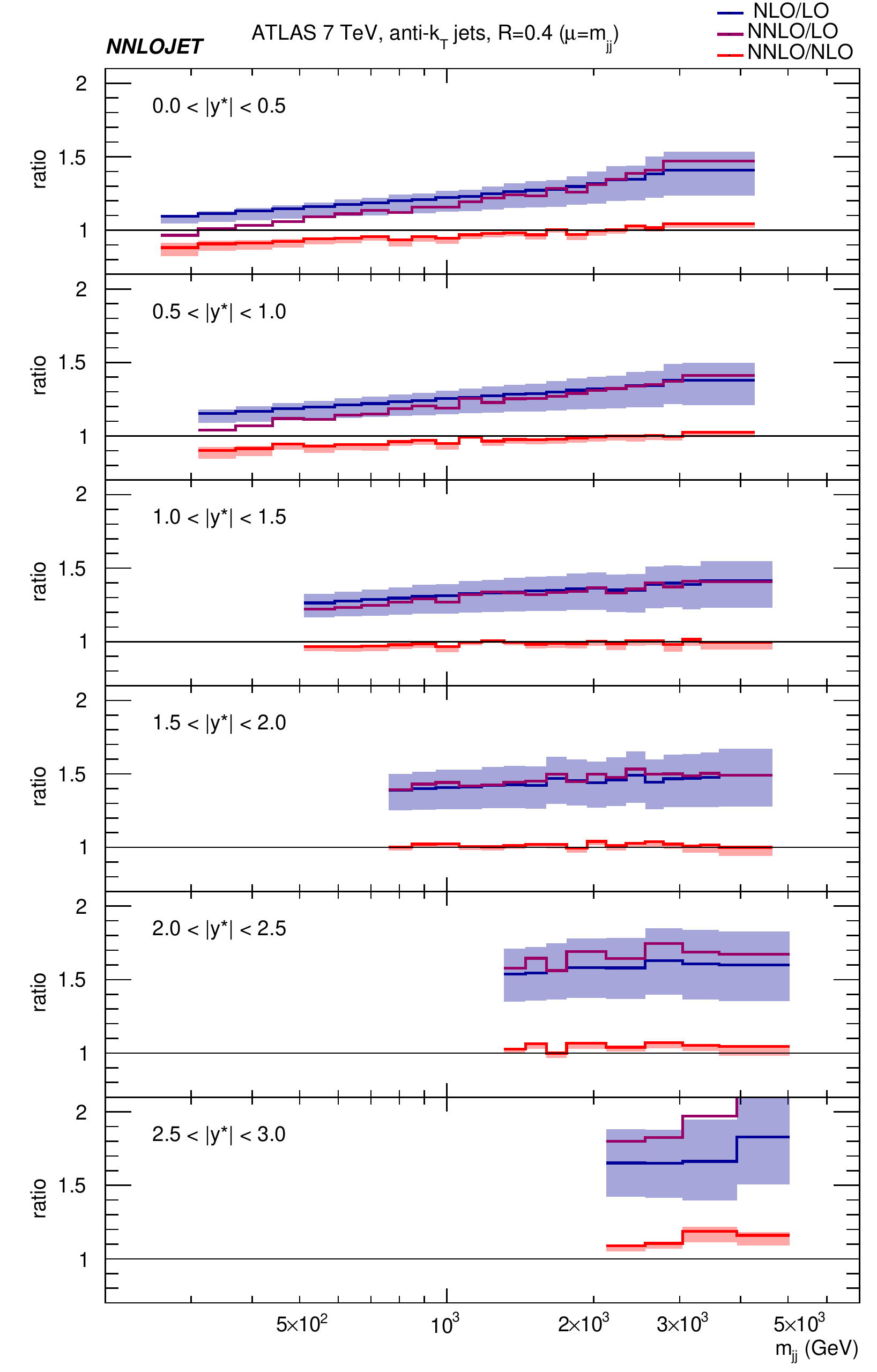}
  \caption{NLO/LO (blue), NNLO/NLO (red) and NNLO/LO (purple) $K$-factors double differential in $m_{jj}$ and \ys. Bands represent
  the scale variation of the numerator. NNLO PDFs are used for all predictions.}
  \label{fig:kfac}
\end{figure}

  \begin{figure}[t]
  \centering
    \includegraphics[width=0.45\textwidth]{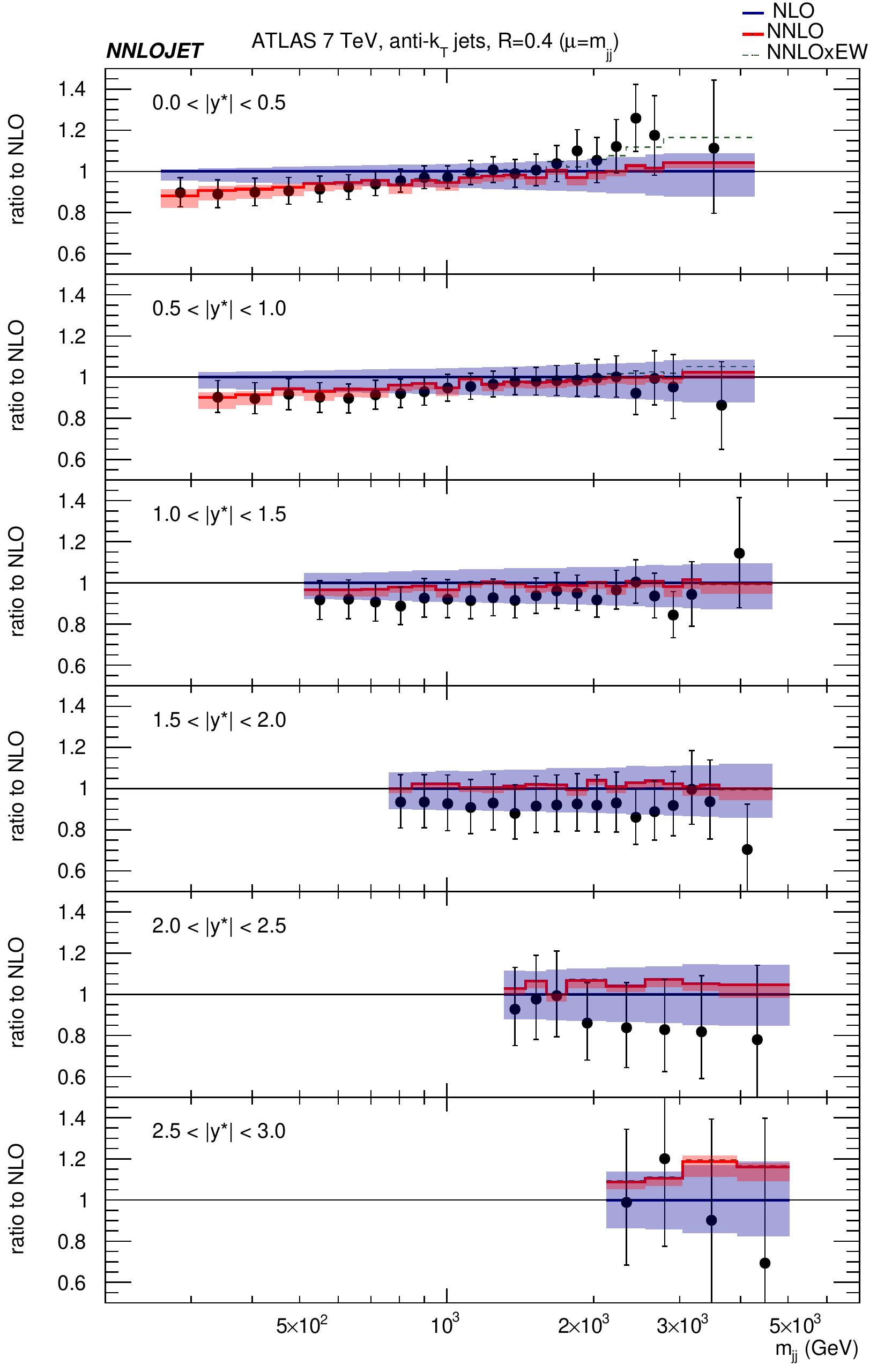}
  \caption{The NLO (blue) and NNLO (red) theory predictions and ATLAS data normalized to the NLO central value. The bands represent
  the variation of the theoretical scales in the numerator by factors of 0.5 and 2. Electroweak effects are implemented as a multiplicative factor and 
  shown separately as the green dashed line.}
  \label{fig:rat2nlo}
\end{figure}

In Fig.~\ref{fig:xsec} we present the absolute cross section as a function of $m_{jj}$ for each \ys bin, compared to NNLO-accurate theory. We observe excellent
agreement with the data across the entire kinematic range in $m_{jj}$ and \ys, with up to seven orders of magnitude variation in the cross section.
The total NNLO prediction shown in Fig.~\ref{fig:xsec} is the sum of LO, NLO and NNLO contributions. We can understand the relative
shift in the theoretical prediction from each perturbative correction by examining the $K$-factors shown in Fig.~\ref{fig:kfac}.
We observe moderate NLO/LO corrections from +10\% at low $m_{jj}$ and \ys to +50-70\% at high $m_{jj}$ and high \ys. The NNLO/NLO
$K$-factors are typically $<10\%$ in magnitude and relatively flat, although they alter the shape of the prediction at low $m_{jj}$ and low \ys. 

To emphasize the size and shape of the NNLO correction, in Fig.~\ref{fig:rat2nlo} we show the distributions normalized to the NLO prediction. 
On the same plot we show the published ATLAS data, also normalized to the NLO theory prediction. We observe good agreement with the NNLO
QCD prediction across the entire dynamical range in $m_{jj}$ and \ys and a significant improvement in the description of the data for low $m_{jj}$ and
\ys, where NLO does not adequately capture the shape nor the normalization. We include the electroweak effects as a multiplicative factor, as calculated
in~\cite{eweak}, and note that in the region where they are non-negligible (\ys$<0.5$, $m_{jj}>2$~TeV) they improve the description of the data.

 \begin{figure*}[t]
  \centering
    \includegraphics[width=0.32\textwidth]{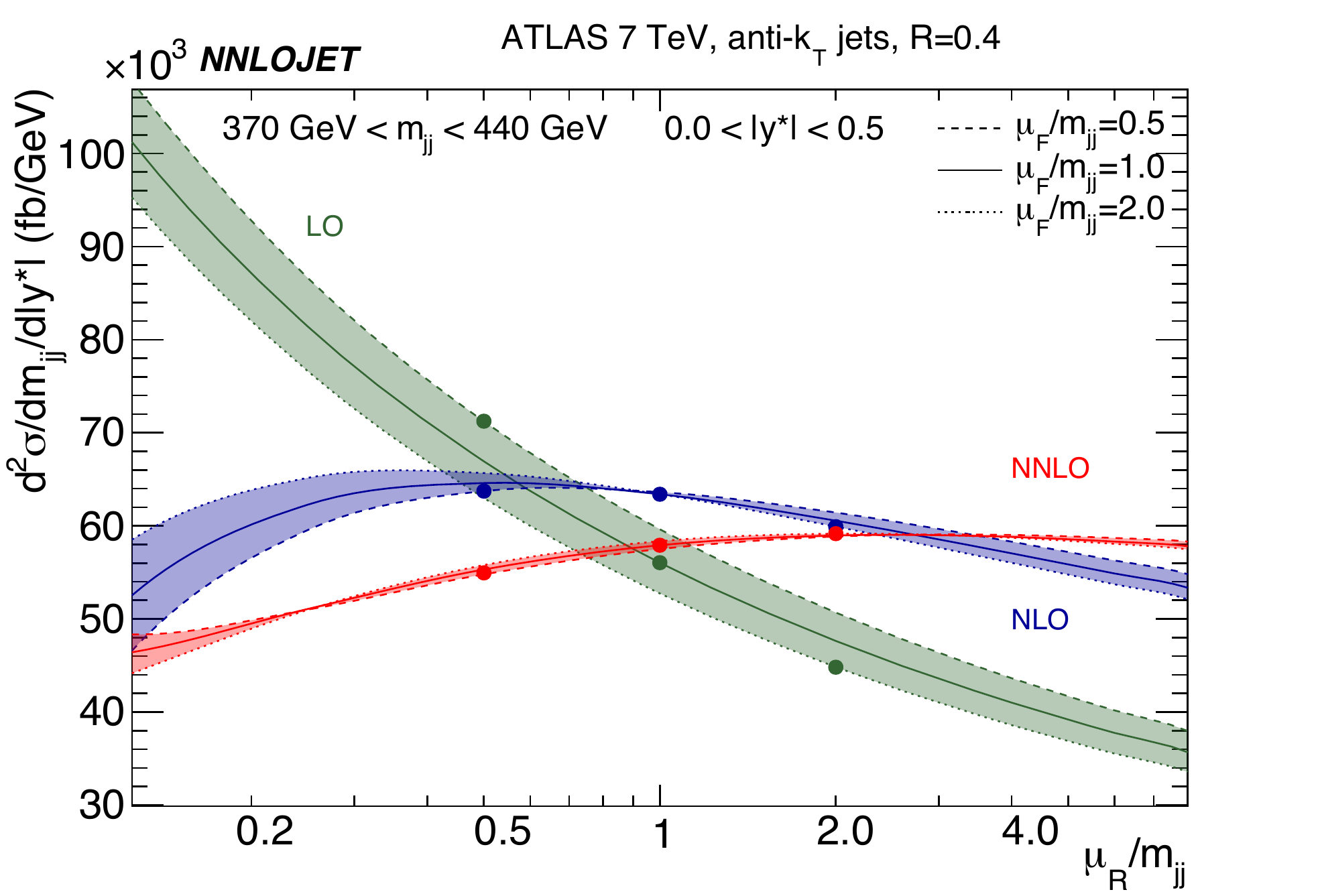}
    \includegraphics[width=0.32\textwidth]{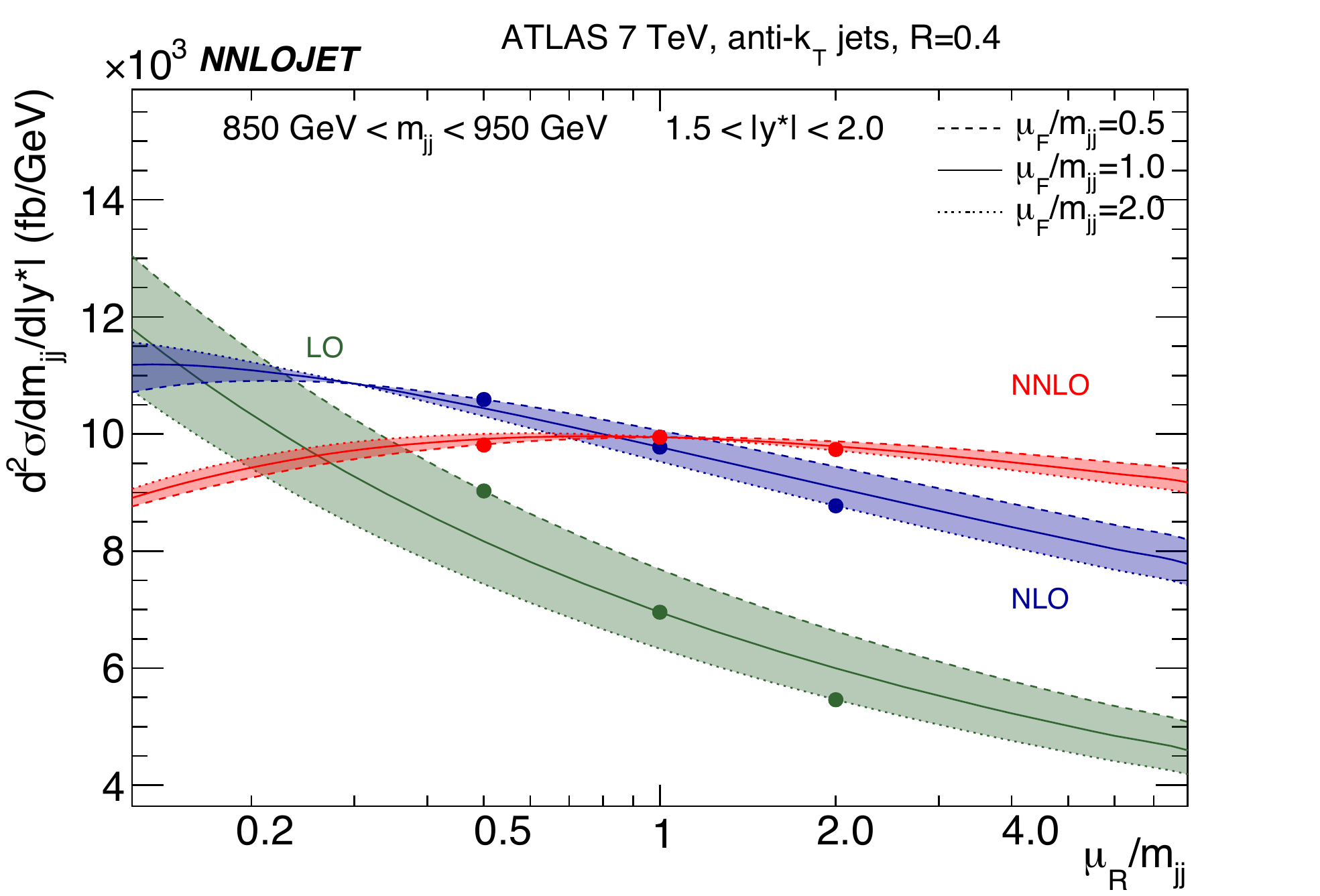}
    \includegraphics[width=0.32\textwidth]{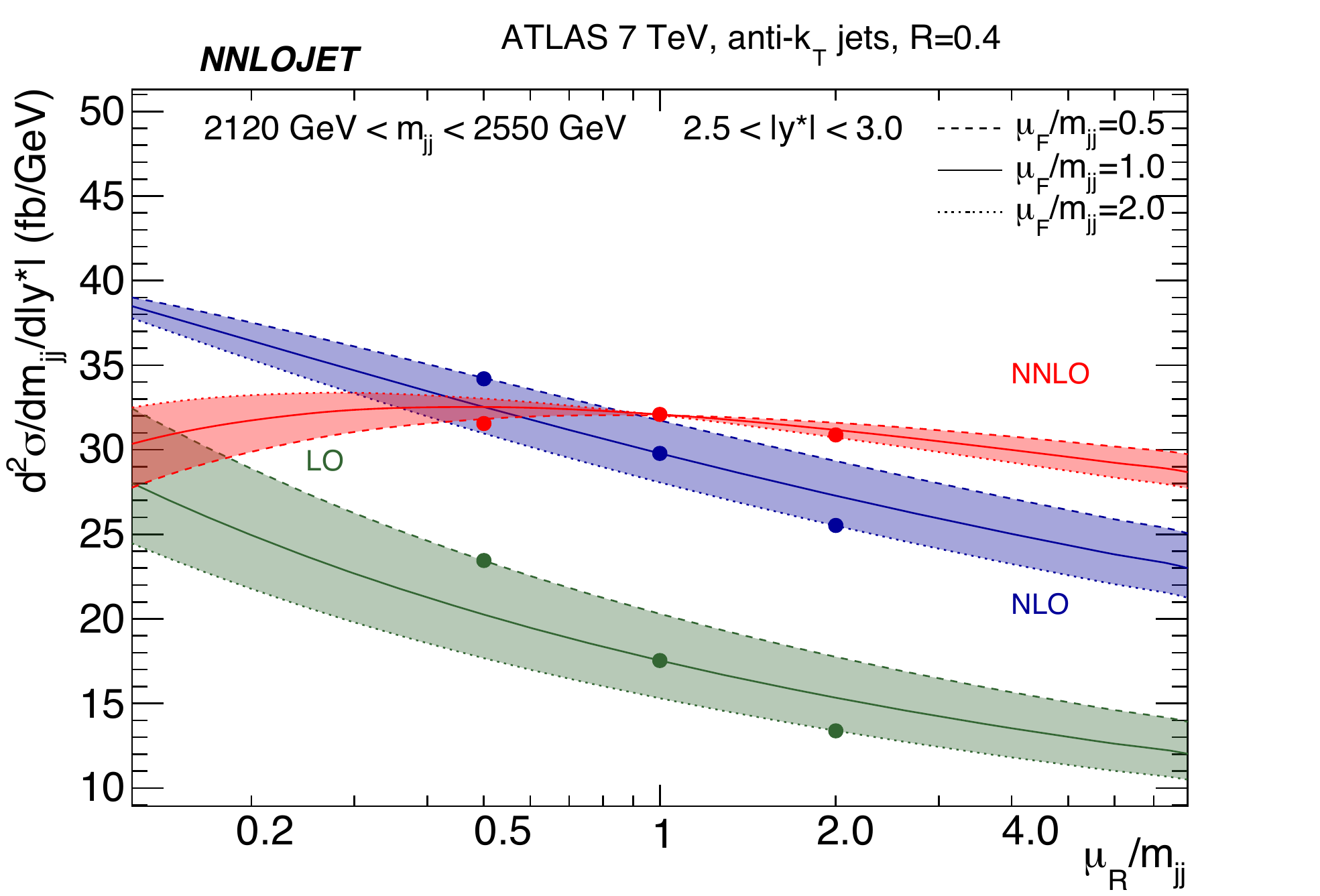}
  \caption{The cross section evaluated in three bins as a function of $\mu_{R}/m_{jj}$: 370~GeV~$<m_{jj}<$~440 GeV, $0.0<|y^*|<0.5$ (left); 850~GeV~$<m_{jj}<$~950 GeV, $1.5<|y^*|<2.0$ (centre); 2120~GeV~$<m_{jj}<$~2550 GeV, $2.5<|y^*|<3.0$ (right). The variation of the cross section, for fixed $\mu_{F}/m_{jj}$, from the central value is shown as solid lines,  computed from the renormalization group equation, for LO (green), NLO (blue) and NNLO (red). The thickness of the bands shows the variation due to factorization
  scale, with boundaries given by $\mu_{F}/m_{jj}=0.5$ (dashed) and $\mu_{F}/m_{jj}=2.0$ (dotted). The points show the \NNLOJET result evaluated at
  $\mu_{R}/m_{jj}=\mu_{F}/m_{jj}=\{0.5,1,2\}$. 
  }
  \label{fig:scalevar}
\end{figure*}

 We generally observe a large reduction in the scale variation and small NNLO
corrections. An exception to this conclusion is found at low $m_{jj}$ and \ys$<1.0$; in this case we observe NNLO scale bands of similar 
size to the NLO bands, and a negative correction of approximately 10\% such that the NNLO and NLO scale bands do not overlap.
To understand this behaviour in more detail we investigate specific bins of $m_{jj}$ and \ys  and study the scale variation inside that bin,
as shown in Fig.~\ref{fig:scalevar}.

The left pane of Fig.~\ref{fig:scalevar} shows the scale variation in the bin 370~GeV~$<m_{jj}<$~440~GeV and $0.0<|y^*|<0.5$, which is the 
region where the NLO and NNLO scale bands do not overlap. For fixed $\mu_{F}$ it is clear that the central scale choice $\mu_{R}= m_{jj}$ lies close
to the extremum of the NLO curve; and the predictions for upper and lower variations of $\mu_F$ cross each other in the vicinity of this central scale. 
As a consequence, the NLO scale variation both in $\mu_R$ and $\mu_F$ is accidentally minimized. The shape of the NLO curve also ensures that the scale variation is asymmetric,
which can be seen in the corresponding bin in Fig.~\ref{fig:rat2nlo}. Notwithstanding the variation in the range $0.5<\mu_{R}/m_{jj}<2$, the NNLO curve
 is clearly flatter and displays less variation than the NLO curve over the full range shown in the left
pane of Fig.~\ref{fig:scalevar}. This suggests that the non-overlapping NLO and NNLO scale bands in this bin is due to the NLO band underestimating the
theoretical uncertainty whereas the NNLO band provides a more reliable estimate. The centre and right panes of Fig.~\ref{fig:scalevar}
show the same quantities in bins of larger $m_{jj}$ and \ys and we see that in these bins the central scale choice $\mu_{R}= m_{jj}$
does not lie near the extremum of the NLO curve, and far away from a crossover point, so we obtain a more reliable NLO scale variation. We
see that the NNLO curves are once again flatter and so we obtain a significant reduction in the scale variation with overlapping NLO and NNLO
scale bands.

In summary, we have presented the first calculation of dijet production doubly differential in $m_{jj}$ and \ys at NNLO and compared to the available ATLAS data.
We find that the ambiguities and pathologies of the theory prediction for certain scale choices at NLO, in particular the $\mu=\langle p_{T}\rangle$ scale choice, 
are removed by including the NNLO contribution. We find that the scale choice $\mu=m_{jj}$ provides a nicely convergent perturbative series with significant
reduction in scale variation at each order in perturbation theory. In particular, the NNLO scale uncertainty is smaller than the experimental uncertainty for this observable.
Overall we observe small NNLO effects which are reasonably flat in $m_{jj}$ and excellent agreement with the data, with the only exception being at low $m_{jj}$ and low \ys
where the moderate NNLO correction improves the description of the data. In this region the NLO and NNLO scale bands do not overlap but this can be
accounted for by the NLO scale band underestimating the perturbative theory uncertainty; whereas we expect the NNLO scale band does provide a reliable estimate. 

It is clear from considering the theoretical uncertainty arising from the parameterization of the scale choice, and the scale variation about that central scale,
that we obtain a reliable theoretical prediction for dijet production for the first time at NNLO. In doing so, the calculation reported here clears the way  
for previously unavailable phenomenological studies using dijet data.
   
  The authors thank Xuan Chen, Juan Cruz-Martinez,
  Tom Morgan and Jan Niehues for useful discussions and their many contributions to the \NNLOJET code. 
  We gratefully acknowledge support and resources provided by the Max Planck Computing and Data Facility (MPCDF). 
  JP would like to thank also Stephen Jones for assistance using the computing facilities of the MPCDF.
  The work of EWNG was performed in part at the Aspen Center for Physics, which is supported by National Science Foundation grant PHY-1066293.
   This research was supported in part by the UK Science and Technology Facilities Council,  
   by the Swiss National Science Foundation (SNF) under contracts 200020-162487 and CRSII2-160814, by the Research 
   Executive Agency (REA) of the European Union under the Grant Agreement PITN-GA-2012-316704 (``HiggsTools'')
    and the ERC Advanced Grant MC@NNLO (340983).

\end{document}